\documentclass{aa}

\usepackage{graphics,epsfig}

\newcommand{\kms}{km~s$^{-1}$}
\newcommand{\cm}{cm$^{-2}$}

\newcommand{\noi}{\noindent}

\begin{document}
\title{A new 21-cm absorber identified with an $L \sim L^\star$ galaxy.}
\titlerunning{A new 21-cm absorber at $z \sim 0.437$}
\author{Nissim Kanekar \inst{1}\thanks{nissim@ncra.tifr.res.in},
Ramana M. Athreya \inst{2}\thanks{rathreya@eso.org}, 
Jayaram N. Chengalur\inst{1}\thanks{chengalu@ncra.tifr.res.in}}
\authorrunning{Kanekar, Athreya \& Chengalur}
\institute{National Centre for Radio Astrophysics, 
Post Bag 3, Ganeshkhind, Pune 411 007 \and European Southern Observatory,
Alonso de Cordova 3107,Vitacura, Santiago 19 Chile}
\date{Received mmddyy/ accepted mmddyy}
\offprints{Nissim Kanekar}
\maketitle

\begin{abstract}
 We present Giant Metrewave Radio Telescope (GMRT) observations of 
redshifted 21-cm absorption from the $z=0.437$ metal line absorption system 
towards PKS~1243-072. HI absorption is clearly detected; the absorption 
profile has a velocity spread of $\sim 20$~km/s. Detection of 21-cm 
absorption indicates that the absorber has an HI column density large 
enough to be classified as a damped Lyman-$\alpha$ system. Follow up ground 
based optical imaging and spectroscopy allow us to identify the absorber 
with an $L \sim L^\star$ galaxy at an impact parameter of $\sim 9.8$~kpc 
from the line of sight to the QSO. The absorbing galaxy is unusual in that 
it has bright emission lines. On the basis of the optical spectrum we are 
unable to uniquely classify the galaxy since its emission line ratios lie 
in the transition region between starburst and Seyfert~II type spectra.
\end{abstract}

\section{Introduction}
\label{intro}

Absorption lines seen in the spectra of distant quasars serve as excellent probes 
of intervening systems along the QSO line of sight. Of these, the highest HI column 
density systems, the so-called damped Lyman-$\alpha$ absorbers (DLAs), are of 
particular interest as they form the major repository of neutral gas at high redshifts. 
By studying the evolution of these DLAs, one can observationally determine the evolution 
of neutral gas in the universe. 

        The connection between the evolution of the neutral gas content and the
average star formation rate in galaxies, however, remains unclear. The HI mass
in DLAs at $z \sim 3$ has been found to be comparable to the stellar mass in
galaxies at $z=0$, consistent with the idea that the gas in the absorbers has
been converted into stars in the intervening period (\cite{storrie96}).
As such, this makes DLAs logical candidates for the precursors of modern-day
spiral galaxies (\cite{wolfe88}). However, the deduced gas mass in DLAs also
depends on the assumed cosmological parameters; in fact, both Storrie-Lombardi \& Wolfe
(2000) and Peroux et al. (2001) point out that, in the currently favoured
$\Omega_\Lambda=0.7, \Omega_{\rm M}=0.3$, $H_0 = 65$~\kms~Mpc$^{-1}$ cosmology, the
estimated gas mass in DLAs at high redshift is, in fact, {\it less} than the mass
in stars at $z=0$ (albeit only at the 1~$\sigma$ level). Interestingly, a recent
Hubble Space Telescope survey for DLAs in a sample of MgII absorbers indicates that
the neutral gas content in DLAs at low redshift is comparable to that at high $z$,
and is, in fact, quite consistent with a scenario in which the HI has {\it not} been
converted into stars (\cite{rt2000}). The latter is, of course, a ``biased'' survey,
since the absorbers were pre-selected on the basis of their MgII absorption; the effects
of this bias are unclear. Regardless of the connection between the evolution of the
neutral gas density and star formation, it remains true that the study of DLAs is
currently the only observational means by which one can trace the evolution of
cold neutral gas in the universe.

	Besides the above, the typical size and structure of damped systems has 
also been an issue of much controversy. Proposed models for DLAs range from large, 
rapidly rotating proto-disks (e.g. \cite{prochaska97}) to small, merging sub-galactic 
systems (e.g. \cite{haehnelt98}). Locally, however, 21-cm emission studies indicate 
that spiral galaxies are the predominant contributors to the HI mass (\cite{rao93});
one would thus expect at least the low  redshift DLAs to be primarily spiral disks.
However, one of the puzzling outcomes of optical imaging 
of low-$z$ DLAs is that 
such systems appear to be associated with a wide variety of galaxy types, with 
only a few systems originating in luminous spirals 
(\cite{lebrun97,turnshek00,turnshek01,cohen01,bowen01}). Further, spectroscopic 
studies indicate that DLAs show very weak (if any) evolution in their metallicity 
with redshift and also do not show the expected $\alpha$/Fe enrichment 
pattern expected for spiral galaxies (\cite{pettini99,centurion2000}). Of course, these 
results could well stem from selection biases in present samples of DLAs arising from, 
for example, issues like dust obscuration. Such issues can be addressed by detailed 
imaging and spectroscopic  observations of individual DLAs; such studies are, however, 
only possible at fairly low redshifts.

Unfortunately, there are very few damped systems known at low redshifts
as their identification requires UV spectra from space-based telescopes.
However, all extra-galactic 21-cm absorbers for which the Lyman-$\alpha$
line has also been observed have HI column densities $N_{\rm HI} > 2 \times 10^{20}$
\cm, i.e. are classically damped. A detection of 21-cm absorption
towards a radio-loud quasar can thus be used as a criterion for the
identification of a damped system; this can then be followed up by optical/UV
studies to identify the absorber.

In this letter, we describe a search for 21-cm absorption at $z = 0.436$ towards 
the radio-loud quasar PKS~1243-072, using the Giant Metrewave Radio Telescope (GMRT). 
The quasar emission redshift is $z_{em}=1.286$ (\cite{wilkes83}). Multiple strong 
low ionization absorption lines (MgII~$\lambda{2798}$, FeII~$\lambda{2600}, 
\lambda{2587}, \lambda{2383}$) have been detected at $z_{abs}=0.436$ towards 
the QSO (\cite{wright79}). The $z=0.436$ absorber was part of the MgII-selected 
sample searched for 21-cm absorption by Lane (2000), with the Westerbork Synthesis 
Radio Telescope, and was classified as a candidate 21-cm absorber on the basis of these 
observations. Our fresh GMRT observations have resulted in the confirmed detection 
of 21-cm absorption in this system. As discussed above, this implies that the absorber 
fits the classical definition of a DLA. We have also carried out  R and I band 
imaging studies of the system, as well as  optical spectroscopy, resulting in the 
identification of the absorber with an $L \sim L^\star$ galaxy.

\section{The GMRT observations}
\label{sec:gmrt}

PKS~1243-072 was observed with the GMRT on the 1st and 2nd of January, 2001, 
using the standard 30 station FX correlator as the backend. A total bandwidth 
of 0.5 MHz was used for the observations, sub-divided into 128 channels; this yielded 
a velocity resolution of $\sim 1.2$~\kms~on each run. Only 18 and 14
antennas were available on the 1st and the 2nd respectively, due to various 
debugging and maintenance activity. The varying baseline coverage is, however, 
unimportant since PKS~1243-072 is unresolved by the longest baselines of the GMRT. 
Phase and bandpass calibration was carried out using the strong nearby source 
PKS~1127-145, while the absolute flux scale was determined using the standard 
calibrator 3C147. The total on-source time was 100 minutes in each run.

%

\begin{figure}
\centering
\epsfig{file=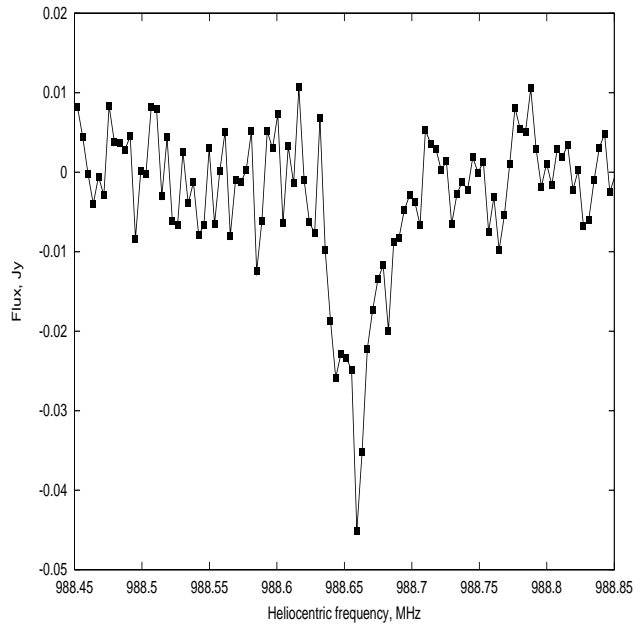,height=3.3truein,width=3.3truein}
\caption{GMRT 0.5 MHz HI spectrum towards PKS~1243-072. The $x$-axis is
heliocentric frequency, in MHz. The spectrum has a resolution of $\sim$ 1.2~\kms.}
\label{fig:gmrt}
\end{figure}

The data were converted from the telescope format to FITS and analysed in AIPS 
using standard procedures. Data from the two different days were analysed separately. 
Continuum emission was subtracted by fitting a linear polynomial to the U-V 
visibilities, using the AIPS task UVLIN. The continuum-subtracted data were then 
mapped in all channels and spectra extracted at the quasar location from the resulting
three-dimensional data cube. The spectra of the two epochs were corrected to the 
heliocentric frame outside AIPS and then averaged together. Finally, the flux of 
PKS~1243-072 was measured to be 480 mJy at both epochs. Our experience with the GMRT 
indicates that the flux calibration is reliable to $\sim 15$\%, in this observing mode.

The final GMRT spectrum of the $z \sim 0.437$ system is shown in Fig.~\ref{fig:gmrt}. 
The RMS noise on the spectrum is 5 mJy, per 1.2~\kms~channel. The absorption can be 
seen to be spread over $\sim$~20~\kms, with a peak optical depth of $\tau = 0.094$.
This occurs at a heliocentric frequency of 988.659 MHz, corresponding to a redshift of 
$z = 0.436699 \pm 0.000003$.

\section{Optical imaging and spectroscopy}

We first imaged the quasar field in the R- and I-bands to identify 
candidates responsible for the absorption lines seen at 21-cm (this paper) 
and optical (\cite{wright79}). As discussed below, the optical imaging revealed 
the presence of a bright galaxy close to the line of sight to the QSO. Subsequent 
long slit spectroscopy of the neighbour showed that this system is indeed at 
$z = 0.437$ and hence likely to be the 21-cm absorber.

\subsection{Optical imaging}

\begin{figure}
\centering
\epsfig{file=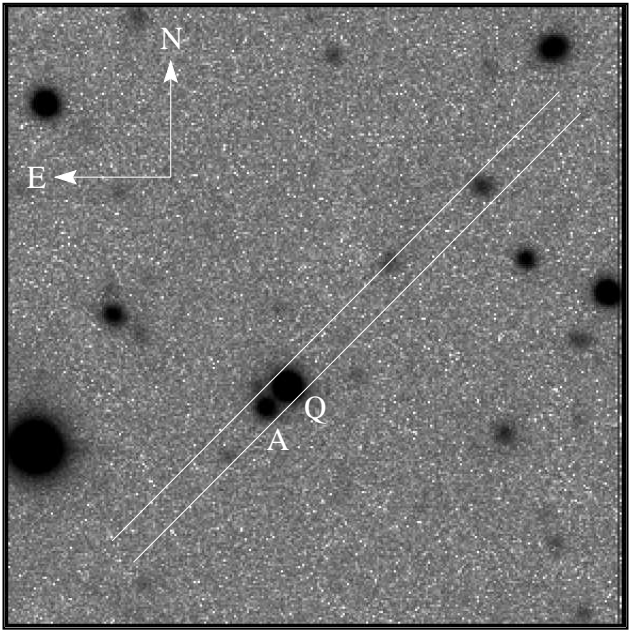,height=3.3truein,width=3.3truein}
\caption{An R-band image of the field of the quasar PKS $1243-072$. The grey-scale
plot is 45 arcsec on each side with the orientation as marked. The two
lines represent the long-slit (PA = $-44.9$~deg) used to obtain the spectra
of the quasar (Q) and the candidate absorber (A). The 2 objects are
separated by 2.2 arcsec on the sky.}
 \label{fig:image}
\end{figure}

The quasar field was imaged using the EFOSC2 instrument on the ESO 3.6m
telescope at La Silla during the night of 29th January, 2001. The Bessel R-band
image was obtained on CCD \#40 in the unbinned mode (0.157 arcsec/pix, 2060 $\times$
2060 pixels $\equiv$ 5.4 $\times$ 5.4 arcmin field). We obtained a total of ten
dithered images of 450s each. The calibration observations included
twilight sky flats and observations of the standard field RU152 (\cite{landolt92}).
We used the science frames to construct and subtract the fringe pattern from the
frames. The atmospheric extinction and the CCD colour term were calculated
using observations of standard fields taken at several airmasses and in
other bands. We expect the photometry to be accurate to about 5\% and use a
conservative error of 0.1 mag. The images were reduced, calibrated, registered
and co-added in a standard manner using IRAF. The final image obtained had a
zero-point of R=32.601 with an RMS of 10.5 counts/pix.

Figure \ref{fig:image} shows a greyscale plot of a 45 $\times$ 45 arcsec section of 
the field around 
the quasar. The quasar (Q) and a rather bright neighbour (A), separated by 2.2 arcsec, are 
visible, close to (and south of) the centre of the field. The quasar
magnitude was measured to be $R = 19.55$ while that of its neighbour was found to be
 $R = 21.27$.

We also obtained an image of the field using the Magellan 6.5m telescope at the Las
Campanas Observatory. The Harris I-band image was obtained on a direct CCD
camera in the unbinned mode (0.069 arcsec/pix, 2048 $\times$ 2048 pixels $\equiv$ 2.36
$\times$ 2.36 arcmin field). We obtained 4 exposures of 600s each; the images were
reduced in the same manner as before. However, it may be noted that, due to technical
problems, we could not obtain a sky/dome flat and had to use the dark-sky flat
to flat-field the image. The fringes on the images were hence divided out rather
than subtracted which resulted in an additional error of a few per cent in the
photometry across the frames. However, the total photometric error is again
about 0.1 mag.

We measured magnitudes of $I = 19.41$ for the quasar and $I = 21.27$ for the
neighbour.

\subsection{Optical spectroscopy}

\begin{figure}
\centering
\epsfig{file=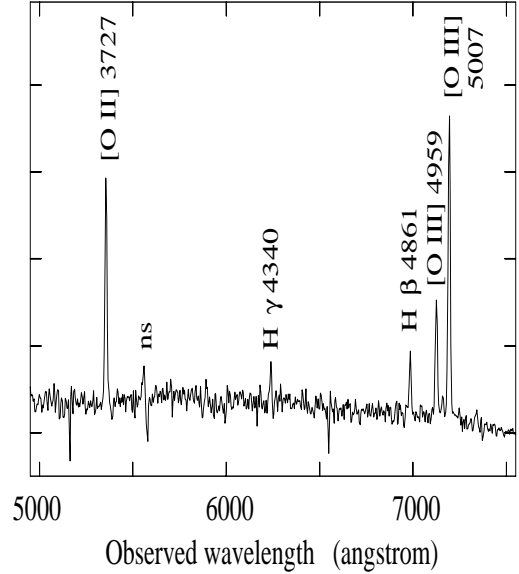,height=3.3truein,width=3.3truein}
\caption{An optical spectrum of the candidate absorber (object~A on Fig. \ref{fig:image}
in the field of the quasar PKS~1243-072. The intensity scale (y-axis)
is linear but in arbitrary units. The labels corresponding to each
emission line identify the species and the rest wavelength (ns : night
sky feature). The average redshift derived from the emission lines is $z = 0.437$.}
\label{fig:spectrum}
\end{figure}

The optical long-slit spectrum was obtained using the EFOSC2 instrument on the
ESO 3.6m telescope during the night of 19th March, 2001. Grism \#11 (300~l/mm,
3300 -- 7520\AA) was used with CCD \#40 (binned 2$\times$2 pixel size 0.314 arcsec) 
to obtain a spectrum with a resolution of 4.1~\AA/pixel. The standard calibration
included He-Ar arc lamp exposures for wavelength calibration and dome flat
exposures to eliminate the pixel-to-pixel gain variation. A 2'' slit was oriented
so as to include both the quasar and the candidate absorber (see Fig. \ref{fig:image}). 
The observations were split into 3 runs of 1800s each, to eliminate cosmic rays.
The data were reduced in a standard manner using the IRAF package.

The spectrum of the candidate absorber is shown in Fig. \ref{fig:spectrum}. It should 
be noted that the night was not photometric and no effort has been made to correct the
spectrum for the CCD spectral response using a standard star observation. However,
we used a previous determination of the CCD spectral response to calibrate the spectra
with the limited aim of estimating spectral line ratios. Thus, while the absolute 
calibration may be significantly wrong, the line ratios should be good to $\sim 15\%$.
We obtained a mean redshift of $z = 0.4370 \pm 0.0001$ using the 5 emission lines
identified in the spectrum. The close correspondence between this emission line
redshift and the redshifts of the absorption lines seen in the quasar spectra
imply that object~A is indeed likely to be the galaxy responsible for the 
absorption line system.

\section{Discussion}
\subsection{HI Column Density}

The 21~cm optical depth, $\tau_{21}$, of an optically thin, homogeneous cloud is related
to the column density of the absorbing gas $N_{\rm HI}$ and the spin temperature $T_{\rm s}$ by
the expression (e.g. \cite{rohlfs86})
\begin{equation}
\label{eqn:tspin}
N_{\rm HI} = { 1.823\times10^{18} T_{\rm s} \over f} \int \tau_{21} \mathrm{d} V \; ,
\end{equation}
\noi where $f$ is the covering factor of the absorber. In the above equation, $N_{\rm HI} $
is in cm$^{-2}$, $T_{\rm s}$ in K and $\mathrm{d}V$ in km s$^{-1}$. For a multi-phase absorber
the spin temperature derived using the above expression is the column density weighted
harmonic mean of the spin temperatures of the individual phases.

In the case of PKS~1243-072, VLBA maps at 2.3 and 8.4 GHz (see the Radio Reference Frame
Image Database of the United States Naval Observatory) show that the entire flux
is contained within $\sim$~20 milliarcseconds. The extremely small size of the background
source makes it highly likely that the covering factor $f$ is close to unity. This implies
an HI column density $N_{\rm HI} = 3 \times 10^{20} (T_{\rm s}/200 {\rm K})$~\cm; given that
all known DLAs have $T_{\rm s} \ge 150$~K (\cite{chengalur00}), it is quite likely that the 
absorber, it is quite likely that the absorber has $N_{\rm HI} > 2 \times 10^{20}$~\cm 
(i.e.  fits the classical definition of a DLA) even if it has a low spin temperature. 
Of course, a spin temperature $\ga 1000$~K, more typical of DLAs (\cite{chengalur00}),
would imply a much higher column density, $N_{\rm HI} \ga 10^{21}$~\cm. However, the 
high luminosity of the absorber makes it likely that its spin temperature is low; this
is discussed in more detail in the next section. Observations of the Lyman-$\alpha$ 
line using the Hubble Space Telescope (HST) will provide a direct estimate of the 
column density and thus, of the spin temperature. 

\subsection{The nature of the Absorber}

	The rather strong optical emission lines seen in the absorber are rather 
unusual. Further, the absorber is only slightly more extended than the 
quasar image (Q$_{\rm FWHM}$ = 0.96 arcsec and A$_{\rm FWHM}$ = 1.08 arcsec 
for the QSO and object~A on the R-band image, and 0.71 and 0.86~arcsec 
respectively, on the I-band image). In fact, object~A appears more 
{\em point-like} than most of the other galaxies in the field. 

	The emission lines could be either due to star-burst activity
or due to the presence of an active galactic nucleus (AGN). It is possible 
to distinguish between these two scenarios on the basis of the ratios of 
the strengths of certain emission lines. Dessauges-Zavadsky et al. 
(2001; hereafter DZ01) present template spectra of different emission 
line galaxies and also discuss their own and earlier classification schemes 
(\cite{rola97,tresse96}). A visual inspection of the template spectra 
clearly shows that the object~A is either a Seyfert~II system or 
a starburst galaxy. Next, the spectrum of object~A has the following 
line ratios 
$$ [O III] \lambda{5007} / H\beta \sim 6.2 $$
$$ [O II] \lambda{3727} / H\beta  \sim 3   $$
$$ [O III] \lambda{4959} / H\beta \sim 2.2 $$

\noindent which can be used as diagnostics for the purpose of classification 
(\cite{rola97,tresse96}). A comparison of these values with Fig. 7 of DZ01 
shows that object~A falls in between the range of ratios obtained in typical 
HII galaxies and Seyfert~II type systems. DZ01 also presented a new diagnostic, based on 
a comparison between the quantity $R_{23} \equiv ([O II] \lambda{3727} + 
[O III] \lambda{4959} + [O III] \lambda{5007})/H_\beta$ and the ratio 
$[OI] \lambda{6300}/H\alpha$. We unfortunately do not have a measurement of 
either $[OI] \lambda{6300}$ or $H\alpha$ in object~A; however, we do estimate 
$\log{[R_{23}]}= 1.06$ for this system. DZ01 noted that 87~\% of all Seyfert~II 
galaxies had $\log{[R_{23}]} > 1.1$ and, an inspection of Fig. 8 in DZ01 
shows that all Seyfert~II systems lie above $\log{[R_{23}]} = 1.05$. 
Again, object~A lies close to the border separating Seyfert~II and starburst 
galaxies. We note that all the above diagnostics tend to place object~A marginally 
amongst the starburst systems. Given the lack of a measured $[OI]/H\alpha$ ratio 
and the problem with spectrophotmetric calibration mentioned earlier, we are
unable to conclusively distinguish between the two possibilities. We plan to carry
out observations of the $[OI] \lambda{6300}$ and $H\alpha$ lines  from this system, 
which should be useful in resolving the issue.

The 2.2 arcsec separation between the quasar and the absorber corresponds to a
linear separation of 9.8 kpc between their lines of sight at the redshift 
of the absorber (assuming a flat FRW Universe, with $H_0
 = 75$~\kms~Mpc$^{-1}$). Although we cannot rule out the  possibility 
that the absorbing galaxy is not object~A, but some fainter companion 
galaxy, the small projected separation between object~A and the QSO 
makes it likely that the absorption arises in object~A itself. We note 
that there is a faint object 1'' north of A and 2'' east of the quasar, 
barely visible in Fig.~\ref{fig:image}. This system is about 2 magnitudes 
fainter than object~A and considerably more diffuse. It is unclear
whether this object is in the vicinity of the QSO or a companion to object~A 
or, indeed, an interloper not associated with either system. 

	Our optical photometry shows that object~A has $L \sim L^\star$. 
This is consistent with the results of Rao \& Briggs (1993) who used a 
survey of the HI content of $z=0$ optically bright galaxies, to conclude 
that the cross section for DLA absorption peaks at this luminosity. Note,
however, that Rosenberg \& Schneider (2001) argue that a substantial 
contribution to the DLA cross-section is provided by optically faint 
galaxies, based on a blind 21-cm survey at $z=0$.
The latter is also consistent with optical searches for the counterparts of 
low redshift DLAs, which have shown that the absorbers arise in galaxies 
with a wide range of luminosities.

Chengalur \& Kanekar (2000) found that low spin temperatures ($T_{\rm s} \la 300$~K) were 
obtained in the few cases where the absorber was identified to be a spiral galaxy; 
such temperatures are typical of the Milky Way and local spirals (see also 
Kanekar \& Chengalur (2001)). However, the majority of DLAs were found to have far 
higher spin temperatures, $T_{\rm s} \ga 1000$~K. Higher $T_{\rm s}$ values are to be expected 
in smaller systems like dwarf galaxies, whose low metallicities and pressures are 
not conducive to the formation of the cold phase of HI (\cite{wolfire95}); such systems 
hence have a higher fraction of warm gas as compared to normal spirals, and therefore, 
a high spin temperature. On the other hand, bright galaxies tend to have high masses, 
and hence both higher metallicities and central pressures, contributing to the 
formation of the cold phase of neutral hydrogen. The high luminosity of object~A thus
indicates that it is likely to have a low spin temperature ($\la 300$~K) and hence, 
a relatively low column density $N_{\rm HI} \la 5 \times 10^{20}$~\cm. It would be 
interesting to test this conjecture by means of HST observations in the Lyman-$\alpha$ 
line, as well as to directly determine the metallicity of the absorber through high 
resolution absorption studies. In a subsequent paper, we plan to compare the metallicity 
as computed from such a high resolution absorption spectrum to the metallicity 
measured from the emission lines.

\begin{acknowledgement}
The radio observations presented in this paper would not have been possible 
without the many years of dedicated effort put in by the GMRT staff to build 
the telescope. The GMRT is run by the National Centre for Radio Astrophysics 
of the Tata Institute of Fundamental Research. The optical observations were
carried out using European Southern Observatory facilities at La Silla 
Observatory, Chile. This research has made use of the United States Naval 
Observatory  (USNO) Radio Reference Frame Image Database (RRFID). This research
has made use  of the NASA/IPAC Extragalactic Database (NED) which is operated 
by the Jet Propulsion Laboratory, California Institute of Technology, under 
contract with the National Aeronautics and Space Administration. 
\end{acknowledgement}
{\bf Note added in proof:} 21cm absorption from 1243-072 has recently been
independently detected by Lane et al. (2001) using the WSRT.

\end{document}